\documentclass{camera}
\usepackage{graphicx}  

\begin{document}

%
\title{TEV ASTROPHYSICS, A REVIEW}

%
\author{Wei Cui}

%
\organization{Department of Physics, Purdue University\\ West Lafayette,
IN 47907, USA}

\maketitle

\begin{abstract}
In this paper I briefly review recent progress in the field of ground-based 
gamma ray astrophysics.
\end{abstract}

\vspace{1.0cm}

%
\section{Scope}
It is not my intention to review the historical development of ground-based 
gamma ray astrophysics or the experimental techniques employed in the field. 
Interested readers are referred to an excellent recent review on those 
subjects~\cite{Weekes06}. 
Neither is it my intention to discuss every source that has been detected,
because this is no longer practical. Instead, I will present a personal 
(thus biased and incomplete) perspective of the scientific impact of TeV 
observations on astronomy and physics. 

\section{Overview}
Since the detection of TeV photons from the Crab Nebula~\cite{Weekes89}, 
the field of ground-based gamma ray astrophysics has experienced steady, 
albeit slow at times, growth in the 1990s and early 2000s. The sources 
detected were predominantly blazars, a subclass of active galactic nuclei 
(AGN), although several other important classes of sources, including pulsar
wind nebula (PWN), shell-type supernova remnant (SNR), and radio galaxy,
were also represented. Most intriguing was arguably the discovery of a source 
of TeV radiation in the Cygnus region\cite{Neshpor95,J2032,Lang04}, which, 
to date, still cannot be associated with any 
known astronomical systems. The source, designated as TeV J2032+4130, is 
therefore the first unidentified TeV source. 

The growth of the field has accelerated drastically since 2003, thanks to the 
availability of a new generation of TeV observatories (HESS in particular; 
see~\cite{Hoffman} for a review on recent HESS results), not only in 
terms of the total number of new sources seen but also in the significant 
increase of the number of Galactic and unidentified sources detected. Table~1 
provides an updated list of the established TeV sources, as of July 2006. 
Figure~1 shows the distribution of these sources in the sky. The clustering 
of unidentified sources around the Galactic plane is interesting but, at 
least to some extent, is the result of a strong observational bias. In 
addition to discrete sources, diffuse TeV emission is detected near the 
Galactic Ridge~\cite{GC_ridge} and in the Cygnus region~\cite{Cygnus}. 

It is perhaps instructive to compare the development of the field with that of
X-ray astronomy. I think that the field is roughly where X-ray astronomy was 
during the {\it Einstein} days, when direct imaging became possible for the 
first time. Coincidentally, it took both fields about the same amount of time 
to reach such a stage, following the detection of the first astronomical 
object (other than the Sun). A noteworthy difference is that a full sky 
survey had already been conducted with {\it Uhuru} before {\it Einstein} was 
launched. Survey is proven to be the most effective way of discovering new
sources and thus providing guidance for deeper follow-up observations. TeV
astrophysics is no exception. Many of the most exciting recent results have
all come out of the HESS survey of the Galactic center 
region~\cite{HGC_1,HGC_2}.
\begin{table}[tp]
\begin{center}
\caption{Known TeV Sources}
{\footnotesize
\begin{tabular}{llllll}
\hline\hline
Class & Name & RA (2000) & Dec (2000) & References$^\dag$ & Notes \\
\hline
Blazar & 1ES 1101-232&11 03 37.57& -23 29 30.2&~\cite{E1101} & \\
 & Mrk 421& 11 04 27.31&    +38 12 31.8& ~\cite{Punch92} & \\
 & Mrk 180&         11 36 26.41&     +70 09 27.3&~\cite{M180} & \\
 & 1ES 1218+304&    12 21 21.94&     +30 10 37.1&~\cite{E1218} & \\
 & H 1426+428 &     14 28 32.6&      +42 40 29&~\cite{H1426}   & \\
 & PG 1553+113&     15 55 43.04&     +11 11 24.4&~\cite{P1553_H,P1553_M}  & \\
 & Mrk 501&         16 53 52.22&     +39 45 36.6&~\cite{Quinn96}  & \\
 & 1ES 1959+650&    19 59 59.85&     +65 08 54.7&~\cite{Holder03,E1959} & \\
 & PKS 2005-489&    20 09 25.39&     $-$48 49 53.7 &~\cite{P2005} & \\
 & PKS 2155-304&    21 58 52.07&     $-$30 13 32.1 &~\cite{P2155}& \\
 & 1ES 2344+514&    23 47 04.92&     +51 42 17.9&~\cite{E2344}  & \\
 & H 2356-309&      23 59 07.8&      $-$30 37 38&~\cite{E1101} & \\
\hline
FR I & M 87&         12 30 49.42&     +12 23 28.0 &~\cite{M87} & \\ 
\hline
Pulsar & Crab Nebula& 05 34 31.97& +22 00 52.1&~\cite{Weekes89}  & PSR B0532+21 \\
Wind &Vela X&          08 33 32   & $-$45 43 42&~\cite{VelaX} & PSR B0833-45\\
Nebula & G313.3+0.1& 14 18 04&	$-$60 58 31 &~\cite{G313} & ``Rabbit'' (R2/Kookaburra) \\
  & K3/Kookaburra &  14 20 09 & $-$60 48 36 &~\cite{G313} & PSR J1420-6048 \\
 &MSH 15-52&       15 14 07   & $-$59 09 27  &~\cite{MSH}& PSR B1509-58; composite\\
 &G18.0-0.7&  18 26 03.0 & $-$13 45 44  &~\cite{G18} & PSR J1826-1334 \\
\hline
Shell-Type & RX J0852.0-4622& 08 52 00 &       $-$46 20 00&~\cite{J0852}&  ``Vela Junior'' \\
Supernova & RX J1713.7-3946& 17 13 00 &       $-$39 45 00&~\cite{J1713}& G347.3-0.5 \\
Remnant & G0.9+0.1&        17 47 23.2&      $-$28 09 06&~\cite{G0.9}& composite\\
 & G12.82-0.02&     18 13 36.6&      $-$17 50 35&~\cite{HGC_1,HGC_2,G12.82} & \\
 & Cas A&           23 23 24 &       +58 48 54&~\cite{casa} & \\
\hline
Microquasar & LS 5039& 18 26 15& $-$14 50 53.6&~\cite{LS} & \\
 &LS I +61303$^\ddag$&   02 40 31.67&  +61 13 45.6 &~\cite{LSI}& \\
\hline
Be Binary &PSR B1259-63&  13 02 47.65& $-$63 50 08.7 &~\cite{B1259}& \\
\hline
Unidentified & HESS J1616-508&  16 16 23.6& $-$50 53 57&~\cite{HGC_1,HGC_2}& PSR J1617-5055 \\
Source I: & HESS J1632-478&  16 32 08.6& $-$47 49 24&~\cite{HGC_2}& IGR J16320-4751 \\
with plausible & HESS J1634-472&  16 34 57.2& $-$47 16 02&~\cite{HGC_2} & G337.2+0.1/IGR J16358-4726\\
counterpart & HESS J1640-465&  16 40 44.2& $-$46 31 44&~\cite{HGC_1,HGC_2} & G338.3-0.0/3EG J1639-4702\\
 & HESS J1713-381&  17 13 58.0& $-$38 11 43&~\cite{HGC_2}& G348.7+0.3\\
 & HESS J1745-290&  17 45 41.3& $-$29 00 22&~\cite{Tsuchiya04,Kosack04,GC,Wang06} & G359.95-0.04/SgrA East/SgrA* \\
 & HESS J1804-216&  18 04 31.6& $-$21 42 03&~\cite{HGC_1,HGC_2}& G8.7-0.1/PSR J1803-2137\\
 & HESS J1834-087&  18 34 46.5& $-$08 45 52&~\cite{HGC_1,HGC_2}& G23.3-0.3 \\
 & HESS J1837-069&  18 37 37.4& $-$06 56 42&~\cite{HGC_1,HGC_2}& G25.5+0.0/AX J1838-0655\\
\hline
Unidentified & HESS J1303-631&  13 03 00.4& $-$63 11 55&~\cite{J1303}& \\
Source II: & HESS J1614-518&  16 14 19.0& $-$51 49 07&~\cite{HGC_1,HGC_2} & \\
with no & HESS J1702-420&  17 02 44.6& $-$42 04 22 &~\cite{HGC_2}& \\
counterpart & HESS J1708-410&  17 08 14.3& $-$41 04 57&~\cite{HGC_2}& \\
 & HESS J1745-303&  17 45 02.2& $-$30 22 14&~\cite{HGC_2} & 3EG J1744-3011\\
 & TeV J2032+4131&  20 31 57& +41 29 56.8&~\cite{Neshpor95,J2032,Lang04}& \\
\hline
Diffuse & Galactic Ridge & --- & --- &~\cite{GC_ridge} & \\
Emission & Cygnus Region & --- & --- &~\cite{Cygnus} & \\
\hline\hline
\end{tabular}
\begin{flushleft}
$^\dag$Including only references in which the detection is first reported 
or confirmed or the counterpart(s) suggested.\\
$^\ddag$Also a Be binary.
\end{flushleft}
}
\end{center}
\end{table}

\section{Origin of TeV Photons}

The precise mechanism for producing the detected TeV photons is not understood
at present. It probably varies from one class of sources to the other, or even 
from one source to the other in the same class. The proposed theoretical
scenarios fall into two broad categories, leptonic models and hadronic models, 
hinging critically on the physical nature of the emitting particles. Although 
details vary for different classes of sources, the general ideas are quite
similar. In leptonic models, the TeV photons are attributed to inverse
Compton (IC) scattering of low-energy photons by TeV electrons (and also 
positrons). The sources of seed photons may include synchrotron radiation
from the electrons themselves (i.e., synchrotron self-Compton, or SSC for 
short), other radiation within the system (e.g., emission from an accretion 
disk, companion star, broad-line regions, etc), and diffuse background 
radiation (e.g., CMB). In hadronic models, TeV emission is thought to be
associated with $\pi^0$ decay and, in some cases (e.g., blazars), with
synchrotron radiation from ultra-relativistic protons in a strong magnetic
field. In the following I briefly summarize the impact of recent results on 
our understanding of emission processes for each class of TeV sources.
\begin{figure}[t!]
\begin{center}
\includegraphics[width=3.5in,angle=90]{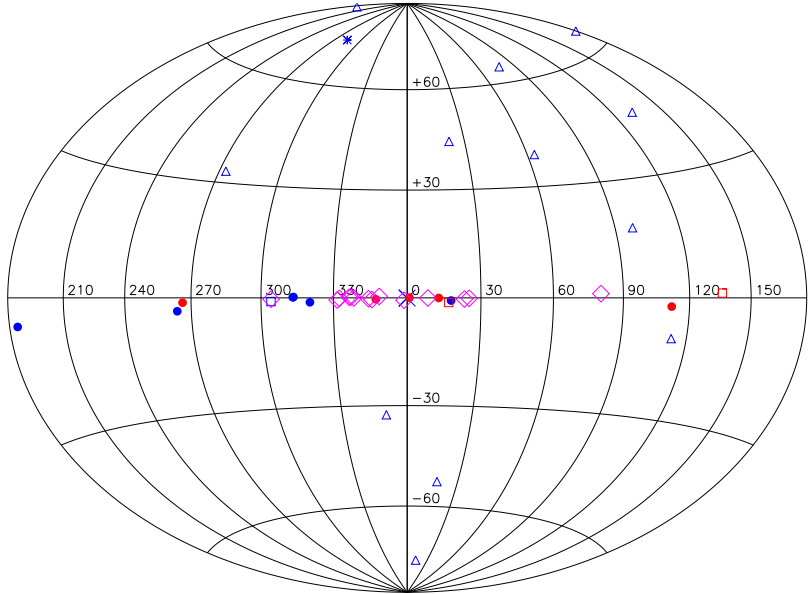}
\label{fig01} 
\caption{Distribution of TeV sources in galactic coordinates. Blazars are in
blue triangles, M 87 in blue asterisk, HESS J1745-290 in blue cross, PWNe in 
blue filled circles, SNRs in red filled circles, microquasars in red squares, 
PSR B1259-63 in blue square, and unidentified sources in magenta diamonds. 
Note that the two PWNe (K3 and Rabbit) in the Kookaburra complex are too 
close to be separated. }
\end{center}
\end{figure}

\subsection{Blazars}

The emission from a blazar is generally thought to be dominated by radiation 
from a relativistic jet that is directed roughly along the line of sight
 (see~\cite{urry95} for a review). The leptonic models associate the TeV 
emission with the SSC process in the jet~\cite{Marscher85,Maraschi92,
Dermer92,Sikora94}. Such models are mainly motivated by the observed 
correlation between X-ray
and TeV variabilities, as well as by the generic ``double-hump'' profile of 
measured SEDs. However, the simple (thus commonly used) version of the
models are increasingly challenged by apparent exceptions to the ``rules'', 
such as relatively loose (or the lack of) X-ray/TeV 
correlation~\cite{Cui04,Blazejow05,PKS2155} and TeV flares that either have no 
counterparts at X-ray energies or that are significantly offset in time 
from their X-ray counterparts~\cite{Kraw04,Cui04,Blazejow05}. The latter are
sometimes referred to as ``orphan TeV flares'', even though they might not 
be truly orphan~\cite{Blazejow05}. To clarify some confusion in recent 
literature, I wish to note that there is hardly any difference between such
flares in 1ES 1959+650 and in Mrk 421, as discussed in ~\cite{Blazejow05}.
Moreover, it has become difficult to fit the measured broadband SED with a 
one-zone SSC model (e.g.,~\cite{Blazejow05}), now that the quality of data 
is much improved, especially at TeV energies. Additional zones are almost 
certainly needed to overcome the difficulties.

The hadronic processes are more complicated. It could involve photon-initiative
cascade~\cite{Manheim92} or $pp$ collisions~\cite{Dar97,Beall99,Pohl00}, but 
it all boils down to $\pi^0$ decay for gamma-ray production. For TeV blazars,
however, it has been argued that $\pi^0$ decay is not 
important~\cite{Ahaproton00,mucke03}; it is the synchrotron radiation from
ultra-relativistic protons in a strong magnetic field that is responsible
for the observed TeV photons from the blazars. The hadronic models also 
seem capable of describing the broadband SED of blazars~\cite{PKS2155} and,
in principle, of accounting for the X-ray/TeV correlation, if an appreciable
amount of X-ray emission comes from synchrotron radiation of {\it secondary} 
electrons. There is quite some flexibility in this case, because the 
co-accelerated, primary electrons might also contribute to the X-ray band
in a significant manner. This could explain the scatters in the X-ray/TeV
correlation. However, like the leptonic models, the hadronic models also
face the challenge of explaining the observed TeV flares with no or offset 
X-ray counterparts~\cite{Blazejow05}. The most severe challenge is, however, 
provided by the
rapid TeV variability of blazars. In my view, the models are on the brink 
of being ruled out by the observed variability on a timescale of minutes 
both at TeV energies (e.g.,~\cite{Gaidos96}) and X-ray energies~\cite{Cui04a,
Xue05}, at least for some sources.

\subsection{Radio Galaxies}

M~87 is the only source in this class that has been detected at TeV energies.
It is classified as an FR I object. For a long time, it is thought to be a
source of ultra-high-energy particles~\cite{Ginsburg64}. The TeV emission has
been modeled both in the leptonic and hadronic scenarios. In the leptonic
model~\cite{Bai01,Stawarz06}, the TeV photons are thought to originate in the 
IC process, as for blazars, but the jet is quite far from the line of sight.
Consequently, M~87 is thought to be a ``mis-aligned'' blazar. In the hadronic
model~\cite{Protheroe02}, on the other hand, the TeV emission is thought to 
be associated with 
synchrotron radiation from protons, also similar to blazars. Both models
can account for the observed TeV flux. Unfortunately, M~87 is detected at a
relatively low significance level ($\sim 4\sigma$)~\cite{M87}, so little
spectral information is available. The situation is expected to improve
soon with the new generation of TeV observatories.

\subsection{Supernova Remnants}

Perhaps not too surprisingly, most recently-detected Galactic TeV sources
fall in this class. After all, supernova remnants have long been suspected
to be the main source of Galactic cosmic rays. Here, I arbitrarily draw a
line between pulsar wind nebulae and shell-type supernova remnants. To 
date, six PWNe and five SNRs are fairly well established as sources of
TeV photons (see Table~1). 

The TeV emission mechanism is thought to be similar for PWNe and SNRs in 
leptonic scenarios but the detailed processes of particle acceleration are 
expected to be very different. The TeV photons are produced in the IC 
scattering of mainly CMB photons by relativistic electrons accelerated 
either by a pulsar~\cite{MSH,G18,VelaX,G313} or by a shock front at the 
outer edge of an SNR~\cite{J1713a,G0.9,G12.82}, although other sources of
seed photons (e.g., star light) may be needed in some cases to account for 
the observed broadband SED. Since the X-ray photons are attributed to
synchrotron radiation from the same electrons, the leptonic models can 
naturally explain the observed spatial coincidence between the X-ray and
TeV emitting regions. Quantitatively, however, such models still have to
overcome some difficulties in fitting the shape of the SED~\cite{J1713a}.

On the other hand, the observed TeV emission from PWNe and SNRs can also be 
explained as the product of the decay of neutral pions that are produced in 
the collision between the relativistic protons and surrounding 
medium~\cite{J1713a,Horns06}.
In this case, the challenge is to account for the observed X-ray/TeV spatial 
correlation. One can either attribute X-rays to synchrotron radiation from 
{\it co-accelerated} electrons~\cite{J1713a,Horns06} or invoke correlated 
enhancement of the magnetic field and the density of the surrounding 
medium~\cite{J1713a}. At
present, it is fair to say that no conclusive evidence exists for the
acceleration of protons in SNRs.

\subsection{Microquasars}

Both TeV microquasars, LS~5039 and LS~I~+61 303, fall in the category of
high-mass X-ray binaries (HXMBs). This is intriguing, because this class
of sources is dominated by low-mass X-ray binaries (LMXBs). There are 
a number of noteworthy differences between the LMXB and HMXB microquasars.
First of all, the luminous high-mass stars in the HMXBs provide a strong 
photon field for IC scattering, while such seed photons are 
expected to play a negligible role in the LMXBs. Secondly, the accretion 
process in the LMXBs is almost certainly mediated by the overflowing of 
the Roche-lobe of low-mass companion stars, while wind accretion is 
probably important in the HMXBs. Thirdly, the LMXBs are exclusively 
transient X-ray sources, while the HMXBs are exclusively persistent X-ray 
sources. Finally, the LMXBs (at least some of them) are known to experience 
episodic ejection events, with superluminal motion seen in some cases, 
while the jets appear to be steady, persistent, and only mildly relativistic 
in the HMXBs. I speculate that some of these differences may be relevant to 
the lack of detection of any LMXB microquasars at TeV energies. 

Like other classes of TeV sources, both leptonic~\cite{Bosch06,Dermer06,
Gupta06a,Paredes06,Gupta06b,Bednarek06} (see also~\cite{Dubus06} for a 
different
interpretation) and hadronic~\cite{Romero05a,Romero05b} models have been 
applied to TeV microquasars. In the leptonic models, the TeV emission is 
associated with IC scattering by electrons. Possible sources of seed photons 
include synchrotron photons (i.e., SSC), stellar photons, disk photons, or
``corona'' photons, but the first two appear to be most important. It is, 
however, being debated whether the SSC or stellar IC gives rise to the
TeV emission in microquasars~\cite{Dermer06,Paredes06}. A dominant stellar IC
scenaro would naturally explain the absence of LMXB TeV microquasars, as well
as the observed orbital modulation of TeV emission~\cite{LSI,LSorb}, given 
both LS~5039 and LS~I~+61~303 are in highly eccentric orbits. In the 
context of the SSC scenario, the orbital modulation could still be attributed 
to $\gamma\gamma$ attenuation~\cite{Dermer06,Gupta06b}, if the TeV emission
originates in the innermost region of the jet. The challenge is to explain
why the SSC process apparently ceases to operate in the LMXBs, in spite of 
the fact that the jets in such sources tend to be more relativistic. To make
progress, a systematic observational effort is required not only to increase 
the size of the sample, but to collect {\it simultaneous} multiwavelength data.
To date, modeling efforts have been based on non-simultaneous data, so the
results should be taken with a grain of salt, given that microquasars are
known to be highly variable on a wide range of timescales. 

\subsection{Be X-ray Binaries}

Microquasars are not the only X-ray binaries detected at TeV energies. 
PSR B1259-63 is also an X-ray binary that contains a 48-ms pulsar in a 
highly eccentric orbit (with a period of $\sim$3.4 years) with a Be star.
Typical of Be X-ray binaries is that the wind of the Be star is thought 
to be confined in the equatorial plane, due to rapid stellar rotation,
and thus forms a dense circumstellar disk. Therefore, the neutron star
interacts with the disk twice for each time it goes around the Be star.
Collision between the winds of the pulsar and Be star leads to the 
formation of a strong shock, which is capable of accelerating particles
to relativistic energies and thus makes such sources promising high-energy
emitters~\cite{Tavani97}. 

The observed TeV emission from PSR B1259-63 can be modeled with IC 
scattering of stellar photons from the Be star by relativistic electrons
accelerated at the shock front~\cite{B1259}. In some sense, it combines
some of the key characteristics of PWNe and HMXB microquasars. Given the 
role of the stellar photon field, one would expect strong orbit modulation
of the TeV emission, which appears to be present. It is interesting to note, 
however, that the TeV emission appears to be at a lull at the periastron 
passage~\cite{B1259}, opposite to what one might naively expect. No detailed
hadronic modeling has been performed. Given the presence of a dense disk
around the Be star, the $pp$ process might be quite efficient here, 
producing neutron pions, which then decay to produce the gamma rays 
seen~\cite{Kawachi04}.

\subsection{Unidentified TeV Sources}

Arguably the most significant recent development in the field is the 
discovery of a population of unidentified TeV sources (see Table~1). Of
course, {\it unidentified} does not necessarily imply {\it unidentifiable} 
or the lack of plausible counterparts. Indeed,
some of the initially unidentified sources have subsequently identified.
For instance, HESS J1813-178 is now associated with a shell-type supernova
remnant (G12.8-0.0)~\cite{G12.82}. The presently unidentified sources 
cluster around the Galactic plane (see Fig.~1), indicating that they are
likely of Galactic origin. Looking at the plausible counterparts (as shown in
Table~1), it appears that most of them might be associated with a PWN or SNR, 
which is supported by non-point-like TeV emission regions. 

However, there are six sources that do not even have a plausible counterpart
(see Table 1). They are being referred to as ``dark accelerators''. Proposals
have been made as to the origin of TeV photons from such systems. For
instance, it has been shown that for old SNRs (with age over $10^5$ yrs) the 
ratio of the TeV flux to the X-ray or radio flux can be very 
high~\cite{Yamazaki06}, which makes it difficult to detect them at long 
wavelengths. On the other hand, HESS J1303-631 
is postulated as the remnant of a gamma-ray burst that occurred in the Galaxy 
some tens of thousands of years ago~\cite{Atoyan06}. 
TeV J2032+4130 is in the general region of the Cyg OB2 association. Strong
winds from massive stars may collide and produce strong shocks that are 
capable of accelerating particles to relativistic energies~\cite{Butt03}. 
The dense environment would be what is required for efficient $pp$ processes 
and subsequent $\pi^0$ decay. More work is clearly required, both 
observationally and theoretically, to reveal the true nature of ``dark 
accelerators''.

\section{Cosmology Connection}

TeV photons may interact with infrared photons to produce electron-positron 
pairs and thus be effectively ``absorbed''. This process must be taken into
account in modeling the SED of all TeV sources. Cosmologically, the 
implication is that we cannot look very far, due to the presence of diffuse
infrared background. The flip side is that we may use observations of distant 
objects at TeV energies to constrain the diffuse infrared background, which
has remained a challenge for direct measurements. The
results can have serious cosmological implications, because of the prospect 
of casting light on star formation in the early universe.

The biggest surprise coming out of recent TeV observations is the conclusion
that the universe seems to be much more transparent than what was initially
thought or, equivalently, the infrared background is much less 
intense~\cite{E1101}. The results are based on observations of two blazars
at moderate redshifts. At the wavelengths of $\sim$1--3 $\mu m$, the derived 
upper limit is barely above the level of integrated light from galaxies that 
have been resolved by {\it Hubble}. It remains to be seen whether the results
can be accommodated by any models. 

The results can be enhanced in two ways. 
One is to extend the spectral coverage to higher energies (beyond 1 TeV), 
e.g., by observing brighter blazars, and thus extend the constraints on the
IR background to longer wavelengths; the other is to observe a large sample 
of blazars at a range of redshifts, with a goal of separating intrinsic and 
extrinsic effects. The latter is important, because our present understanding 
of the intrinsic spectral shape of the blazars (or other objects) is still 
quite incomplete.

\section{Cosmic Ray Connection}

Despite decades of intense observational and theoretical efforts, the origin 
of cosmic rays remains a mystery. For cosmic rays below $\sim 10^{15}$ eV, 
it is almost taken for granted that they are associated with SNRs in the
Galaxy. Strong shocks at the outer edge of SNRs are naturally thought of as
the site for particle acceleration. If it is the case, SNRs would be 
promising candidates for TeV experiments. It is reassuring, therefore, that
a number of SNRs have indeed been detected at TeV energies (see Table~1).
Unfortunately, this success has not led to convincing evidence for the
production of cosmic rays in SNRs. As discussed in \S~3.3, it is still not 
possible, at present, to rule in or rule out either leptonic or hadronic 
scenarios, based on the existing data. It is, however, hopeful that the 
quality of data will be continually improved, which may eventually allow us 
to see some unique characteristics of $\pi^0$ decay.

On a different front, the detection of diffuse TeV emission in the Cygnus
region and around the Galactic Ridge has provided direct evidence for 
interactions between cosmic rays and molecular clouds in the 
Galaxy~\cite{Cygnus,GC_ridge}. Spatial correlation between the Galactic Ridge
emission and molecular clouds leaves little room
for an alternative explanation. The measured gamma-ray spectrum indicates
that the spectrum of cosmic rays near the center of the Galaxy is 
significantly harder than that in the solar neighborhood, presumably due 
to propagation effects~\cite{GC_ridge}. Moreover, the density of cosmic
rays seems to be many times the local density. It is argued that the
observations can be explained by the presence of a particle accelerator
near the Galactic center that has been active over the past $10^4$ years. An 
obvious candidate is the supernova remnant Sgr A East, which has about the 
right age. Moreover, Sgr A East is a plausible counterpart of HESS J1745-290 
(see below), whose TeV spectrum has a similar shape to the spectrum of the 
diffuse emission.

\section{Other Connections}

The detection of HESS J1745-290 raises the possibility of ``seeing'' dark 
matter in the vicinity of the Galactic center with TeV 
observations~\cite{Horns05}. The region has long been thought of as the best 
place to search for gamma-ray emission resulted from the annihilation
of dark matter particles~\cite{Berezinsky94,Berg98,Berg01,Cesarini04,Hooper04}. While the prospect of detecting dark matter is exciting, such
an origin is highly unlikely for HESS J1745-290~\cite{GC,Zaharijas06}, based
on the incompatibility between the shape of the measured gamma-ray spectrum 
and that of theoretical expectations. A more plausible explanation is to
associate the TeV emission with an SNR (G359.95-0.04 or Sgr A East) or Sgr A*,
although the former seems more likely, given the apparent lack of variability 
of HESS J1745-290~\cite{Hoffman}. 

M 87 has also been mentioned as a possible source of TeV photons from dark
matter annihilation~\cite{Baltz00}. TeV data of much improved quality are 
expected to be available from MAGIC and/or VERITAS, so a test of the 
hypothesis may become possible soon. Other suggested sources of TeV emission
from dark matter annihilation include central regions of nearby galaxies 
in general, dwarf spheroidal galaxies, clusters of galaxies, and perhaps 
even globular clusters (see, e.g.,~\cite{Colafrancesco06}, and references 
therein), although there has been no positive detection of any such source 
at TeV energies.

\section{Concluding Remarks}

TeV astrophysics has become a viable branch of modern astronomy. It provides
a unique window to the extreme non-thermal side of the universe. Many classes 
of astronomical systems have been detected at TeV energies. The observations 
have not only shed new light on the properties of the systems themselves but 
on the physical processes operating in diverse astronomical settings. For 
instance, taken together, Blazars, microquasars, and gamma-ray bursts (though
none has been detected at TeV energies yet~\cite{Atkins05,MAGICGRB,Horan06}) 
may offer an excellent opportunity
for us to make some tangible comparisons of the processes of particle 
acceleration and interaction in the jets of black holes over a vast range of 
physical scales (from microparsecs to megaparsecs)~\cite{Cui05}. As the 
capability of TeV observatories improves, it is hopeful that more sources in 
the established classes and, more importantly, new classes of sources will 
be detected.

The field is of equally great interest to physicists, because it has made it 
possible to study some of the most important questions in physics at energies 
much beyond the capabilities of present and future particle accelerators.
Independent of theoretical scenarios, TeV observations are capable of
constraining the intrinsic spectrum of emitting particles and thus casting
light on the nature of the particles and on the acceleration mechanisms. 
TeV observations have already begun to have a serious impact on modern 
cosmology. They have also provided insights into such fundamental issues as 
dark matter, primordial black holes, and Lorentz invariance. The constraints
are expected to improve as the quality of data improves. In many cases, 
however, the challenge is to separate astronomical and physical origins of 
the TeV photons detected.

\section{Acknowledgements}

I wish to thank A. Konopelko for useful discussions and comments on the
paper. I gratefully acknowledge financial support from the U.S. Department
of Energy.

\bigskip
\bigskip
\noindent {\bf DISCUSSION}

\bigskip
\noindent {\bf TERESA MONTARULI:} Can we discuss the HESS plot on the diffuse
emission from 200 pc around the HESS source in the Galactic center region? Do
we really need to increase the density of cosmic rays? The harder spectrum 
can be explained by the absence of propagation effects close to the
accelerator.

\bigskip
\noindent {\bf WEI CUI:} The plot shows the TeV emission in the
region after contribution from the detected point-like sources is removed.
Therefore, it is the amount of diffuse emission that requires a higher cosmic 
ray density than the local average. Of course, there is always the 
possibility of residual contribution from weaker point sources that are below
the detection threshold. However, the striking spatial correlation between 
the TeV peaks and the molecular clouds points to a diffuse origin of the
bulk of the emission. As for the required harder particle spectrum, the 
authors have indeed suggested as a plausible explanation the proximity of 
interaction regions to the sources that accelerate the particles.

%
\end{document}